\documentclass[floatfix,aps,amsmath,amssymb,nofootinbib,superscriptaddress,showkeys]{revtex4}

\usepackage{graphicx,color}
\usepackage{amsmath,amssymb,bm}
\usepackage{dcolumn}
\usepackage{comment}

\newcommand{\bea}{\begin{eqnarray}}
\newcommand{\eea}{\end{eqnarray}}
\newcommand{\be}{\begin{equation}}
\newcommand{\ee}{\end{equation}}
\newcommand{\np}{{\bf p}}

\newcommand{\nh}{{\bf h}}

\newcommand{\nq}{{\bf q}}

\newcommand{\neta}{\mbox{\boldmath $\eta$}}

\begin{document}

\title{
Improved Superscaling in Quasielastic Electron Scattering 
with Relativistic Effective Mass}

\author{P.R. Casale} \affiliation{Departamento de
  F\'{\i}sica At\'omica, Molecular y Nuclear \\ and Instituto Carlos I
  de F{\'\i}sica Te\'orica y Computacional \\ Universidad de Granada,
  E-18071 Granada, Spain.}

\author{J.E. Amaro}\email{amaro@ugr.es} \affiliation{Departamento de
  F\'{\i}sica At\'omica, Molecular y Nuclear \\ and Instituto Carlos I
  de F{\'\i}sica Te\'orica y Computacional \\ Universidad de Granada,
  E-18071 Granada, Spain.}

\author{V.L. Martinez-Consentino}\email{victormc@ugr.es} 
  \affiliation{Departamento de
  F\'{\i}sica At\'omica, Molecular y Nuclear \\ and Instituto Carlos I
  de F{\'\i}sica Te\'orica y Computacional \\ Universidad de Granada,
  E-18071 Granada, Spain.}

\author{I. Ruiz Simo}\email{ruizsig@ugr.es} \affiliation{Departamento de
  F\'{\i}sica At\'omica, Molecular y Nuclear \\ and Instituto Carlos I
  de F{\'\i}sica Te\'orica y Computacional \\ Universidad de Granada,
  E-18071 Granada, Spain.}

\date{\today}

\begin{abstract}
 Superscaling in electron scattering from nuclei is
  re-examined paying special attention to the definition of the
  averaged single-nucleon responses.  The validity of the extrapolation
  of nucleon responses in the Fermi gas has been examined, which
  previously lacked a theoretical foundation. To address this issue,
  we introduce new averaged responses with a momentum distribution
  smeared around the Fermi surface, allowing for momenta above the
  Fermi momentum. This approach solves the problem of negativity in
  the extrapolation away from the scaling region and, at the same
  time, validates its use in the scaling analysis.  This work has
  important implications for the interpretation of scaling data
  and contributes to the development of a more complete understanding
  of the scaling approach. 
\end{abstract}

% Keywords
\keywords{
quasielastic electron scattering; 
superscaling analysis;
relativistic mean field;
relativistic effective mass}

\maketitle

\section{Introduction}

In the field of nuclear physics, understanding the behavior of atomic
nuclei under various conditions is of utmost importance. One such
phenomenon is the electromagnetic response of nuclei in electron
scattering experiments \cite{Wes75,Sar93,Ben08,Bof96,Wal04}.  More
recently, neutrino experiments with accelerators have
increased the interest and the need to describe the electro-weak
response of the atomic nucleus \cite{Alv18,Mos16,Kat17,Alv14}.  
Electron and
neutrino scattering processes are closely related, as the
electromagnetic current is linked to the weak isovector current
\cite{Ama20,Ank22,Ama06}.  Therefore, it is of central importance to
describe first the electromagnetic response, as there is an abundance
of experimental data on these reactions \cite{archive,archive2}.

  In this article, we focus on the nuclear quasielastic response in
  electron scattering, and more specifically, on the superscaling
  model \cite{Don99,Don99a}, whose basic theoretical foundations we
  aim to examine.  One-nucleon emission is the most important
  contribution to the inclusive cross section in the quasielastic
  region, centered around $\omega=|Q^2|/2m^*_N$, where $\omega$ is the
  energy transfer, $Q^2=\omega^2-q^2<0$, and $q$ is the momentum
  transfer to a nucleon with relativistic effective mass $m_N^*$
  \cite{Ros80,Ser86,Dre89,Weh93}.

The widely used model of superscaling assumes factorization of the
electron-nucleus scattering cross section, which is proportional to
the average electron-nucleon scattering probability times a
phenomenological scaling function that incorporates the nuclear
structure information \cite{Alb88,Cen97}.  Despite their limitations,
such as neglecting the effects of final-state interactions and
meson-exchange currents, this approach has the potential to provide an
accurate description of the quasielastc electron and neutrino
 scattering data in the
quasielastic peak region with only a few parameters: the Fermi
momentum, $k_F$, and the relativistic effective mass, $m_N^*$ ---or the
nucleon separation energy depending on the particular approach to
superscaling \cite{Ama21}--- as well as the phenomenological scaling function.
Several methods have been employed to extract the phenomenological
scaling function from experimental data. The SuSA (superscaling
approach) model utilizes longitudinal response data \cite{Ama04},
additionally the SuSA-v2 uses 
theoretical input to construct a scaling function in the transverse channel
\cite{Meg16b},
while the more recent SuSAM* (superscaling approach with relativistic
effective mass) model extract the scaling function directly from cross
section data and incorporates medium corrections through the effective
mass of the nucleon \cite{Mar17,Ama18,Rui18}. 
Attempts to extend the formalism to the inelastic region have also been 
made \cite{Ama04,Mai09,Meg16b}.

Despite the success of the phenomenological SuSA and SuSAM* models in
the quasielastic peak, one aspect of the theory that remains
unverified is the choice of the nuclear average of the single-nucleon
response. In most works, the single-nucleon response was averaged over
the relativistic Fermi gas \cite{Alb88}, and then extrapolated by
analytic continuation to the energy transfer region that is prohibited
in the Fermi gas due to the Pauli blocking effect
\cite{Ama04,Meg16b,Ama18}.  While this approach yield good results,
extrapolating a function outside the range of validity is dangerous
and needs a physical justification.

In this article, we investigate the behavior of the single-nucleon
responses when averaged over a Fermi gas and extrapolated outside of
the kinematic range allowed by the Pauli blocking effect. We show that
as we move further away from the scaling region, the extrapolation
loses its physical meaning and yields negative results for the
response, which should be positive. On the other hand, we demonstrate
that using extrapolation in the scaling region is appropriate because
it produces results similar to those obtained by averaging the
response over a nuclear momentum distribution, which does not suffer
from this issue.

Our proposed framework involves a new definition of the single-nucleon
response averaged over momentum space, with a momentum distribution
where the Fermi surface is smeared out instead of using the sharp
Fermi gas distribution. This average therefore has a theoretical
justification, in contrast to the extrapolation approach
\cite{Ama04,Meg16b,Ama18}, and produces results that are similar to
those of the traditional superscaling models. 
With this approach, we have a solid argument
that justifies the choice of the single-nucleon response and does not
suffer from the previous issues.  While 
we will show that the use of the new averaged
single-nucleon or the extrapolated one is indifferent in the scaling
region, this work improves the superscaling formalism from the
theoretical point of view by providing a physical justification for
its use, which strengthens the applicability of such phenomenological
models. 
 Our findings have implications for the SuSA and SuSAM* model,
as well as for other phenomenological models in nuclear physics.

The scheme of the paper is as follows.  In Sect. 2 we present a brief
review of superscaling formalism in the context of the SuSAM*
approach.  In sect. 3 we analyze the averaged single-nucleon
responses, discuss the problems of the extrapolation, and propose a
new definition. 
We give some details on the calculation of the single-nucleon responses
in Appendix A.
In sect 4 we present results of the single-nucleon
cross sections and perform an updated scaling analysis of the $^{12}$C
data using the new definition. In sect. 5 we discuss the results
and finally in sect 6 we present our conclusions.

\section{Review of superscaling formalism}
%------------------------------------------

In this section we will briefly review the theory of the 
relativistic Fermi gas (RFG) response
function and its connection with the theory of superscaling. The
scaling variable $\psi$ was first introduced in ref. \cite{Alb88}. The
scaling formalism was refined in subsequent works
\cite{Cen97,Don99,Don99a} until reaching the most up-to-date version
of the SuSA-v2 model \cite{Ama20}. 

The formalism in this work is an extension of the SuSA 
to the SuSAM* approach ---based on the
equations of nuclear matter interacting with a relativistic mean field
(RMF) \cite{Ros80,Ser86,Dre89,Weh93}.  The RMF model differs from the
RFG mainly in that the nucleons acquire a relativistic effective mass
$m_N^*$.  The on-shell energy with effective mass is defined as
\begin{equation} \label{on-shell}
E = \sqrt{p^2+(m_N^*)^2}.
\end{equation}
In the RMF this is not the total energy of the nucleon, but rather, the
nucleons acquire an additional positive vector energy that partly
cancels the (negative) attraction energy of the scalar field. 
However in this work  we deal with the one-particle one-hole (1p1h)
 response functions 
where the vector  energy  of particles and holes cancel. So the
response only depends on the effective mass and the Fermi momentum.  

\subsection{Electromagnetic response functions}

We consider the inclusive electron scattering process where an
incident electron with energy $\epsilon$ scatters off a nucleus with
scattering angle $\theta$. The final electron energy is
$\epsilon'$. The momentum transfer is $q$ and the energy transfer is
$\omega$, and $Q^2=\omega^2-q^2<0$. 
The cross section in plane-wave Born approximation with one
photon-exchange is written
\begin{equation}  \label{cross}
\frac{d\sigma}{d\Omega d\epsilon'}
= \sigma_{\rm Mott}
(v_L R_L(q,\omega) +  v_T  R_T(q,\omega)),
\end{equation}
were $\Omega$ is the final electron solid angle, 
$\sigma_{\rm Mott}$ is the Mott cross section, 
\begin{equation}
\sigma_{\rm Mott}=
\left(\frac{\alpha\cos\theta/2}{2\epsilon\sin^2\theta/2}\right)^2,
\end{equation} 
$v_L$ and $v_T$ are the kinematic factors 
\begin{equation}
v_L = 
\frac{Q^4}{q^4}, 
\kern 1cm
v_T =  
\tan^2\frac{\theta}{2}-\frac{Q^2}{2q^2},
\end{equation}
and finally, $R_K(q,\omega)$, $K=L,T$, 
are the longitudinal and transverse response functions defined below.

We focus on the description of the nuclear response functions 
 resulting from the interaction of the electron with the one-body 
electromagnetic
current, giving rise to 1p1h excitation of the
Fermi gas. They are defined in a similar way to the usual RFG formalism 
\cite{Ama20},
with the difference that in our case the nucleons have an effective mass
$m^*_N<m_N$.
The hole momentum is $\nh$ with $h<k_F$ and 
on-shell energy $E=\sqrt{h^2+(m_N^*)^2}$.
By momentum conservation, the particle momentum is $\np'=\nh+\nq$ 
with on-shell energy $E'=\sqrt{p'{}^2+(m_N^*)^2}$. Pauli blocking implies
$p'>k_F$.
 The nuclear response functions are then  given by
 \begin{equation}
R_K^{QE}(q,\omega)
= 
 \frac{V}{(2 \pi)^3}
\int
d^3h
\frac{(m^*_N)^2}{EE'}  2w_K  \,
\theta(p'-k_F)
\theta(k_F-h) 
\delta(E'-E-\omega),
\label{rmf}
\end{equation}
where $w_K$ are the single-nucleon responses for the 1p1h excitation
\begin{eqnarray} \label{snresponses}
w_L =  w^{00},
\kern 1cm
w_T  =  w^{11}+w^{22},
\end{eqnarray}
corresponding to the single-nucleon hadronic tensor
\begin{equation} \label{traza}
w^{\mu\nu}=\frac12 \sum_{ss'} (J^{\mu}_{s's})^* J^{\nu}_{s's}
\end{equation}
and $J^\mu$ is the electromagnetic current matrix element
\begin{equation} \label{corriente}
J^{\mu}_{s's}= \overline{u}_{s'}(\np')
\left[F_1\gamma^\mu+i\frac{F_2}{2m_N}\sigma^{\mu\nu}Q_{\nu}
\right]u_s(\nh),
\end{equation}
where $F_1$ and $F_2$, are the Dirac and Pauli form factors of the nucleon. 
Note that we use the current operator in the vacuum, but the spinors correspond to nucleons with  effective mass $m_N^*$. 

To compute the integral (\ref{rmf}), we
use the variables $E, E', \phi$, with Jacobian 
$h^2 dh d\cos\theta= (EE'/q)dEdE'$.  Then the  integral over $E'$ is made using
 the Dirac delta.  This fixes the angle between $\nq$ and $\nh$ to the value
\begin{equation} \label{angulo}
\cos\theta_h= \frac{2E\omega+Q^2}{2hq}, 
\end{equation}
 and the integration over the azimuth angle $\phi$ gives
$2\pi$ by symmetry of the responses when $\nq$ is on the $z$-axis
\cite{Ama20}. We are left with an integral over the initial nucleon
energy
 \begin{equation}
R_K^{QE}(q,\omega)
= 
 \frac{V}{(2 \pi)^3}
 \frac{2\pi m_N^{*3}}{q}
\int_{\epsilon_0}^{\infty}d\epsilon\, n(\epsilon)\, 2w_K(\epsilon,q,\omega),
\label{respuesta}
\end{equation}
where $\epsilon=E/m^*_N$ is the initial nucleon energy in units of
$m_N^*$, and $\epsilon_F=E_F/m_N^*$ is the (relativistic) Fermi energy
in the same units.  Moreover we have introduced the energy
distribution of the Fermi gas $n(\epsilon)=
\theta(\epsilon_F-\epsilon)$.  The lower limit, $\epsilon_0$ of the
integral in Eq. (\ref{respuesta}) corresponds to the minimum energy
for a initial nucleon that absorbs energy $\omega$ and momentum $q$.
It can be written as (see Appendix C of ref. \cite{Ama20})
\begin{equation}
\epsilon_0={\rm Max}
\left\{ 
       \kappa\sqrt{1+\frac{1}{\tau}}-\lambda, \epsilon_F-2\lambda
\right\},
\end{equation}
where we have
introduced the
dimensionless variables 
\begin{eqnarray}
\lambda  = \omega/2m_N^* & 
\kappa   =  q/2m_N^* &
\tau  =  \kappa^2-\lambda^2. 
\end{eqnarray}

\begin{figure}[t]
\centering
\includegraphics[width=17cm,bb=20 600 540 770]{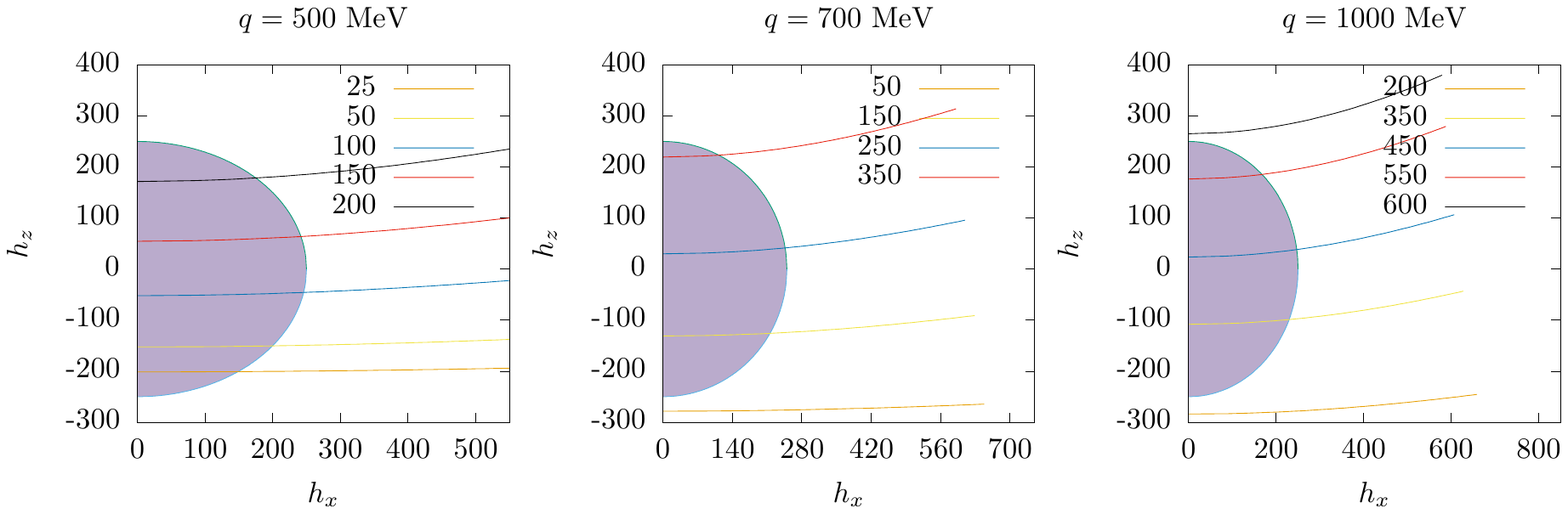}
  \caption{Integration path in momentum space of the initial nucleon
    corresponding to the integral (13) for different values of the
    energy transfer $\omega$ (indicated in MeV in the key for each panel)
 and for three values of the momentum   transfer}
  \label{fig1}
\end{figure}

\subsection{Geometrical interpretation}
%-----------------------------------------

For a fixed value of $\phi, q, \omega$, the integral over energy
$\epsilon$ in Eq. (\ref{respuesta}) corresponds to integrating the
single nucleon response over a path in the momentum space of the hole
$\nh$, weighted with the momentum distribution.  This curve is easily
obtained from Eq. (\ref{angulo}), giving the angle $\theta_h$ as a
function of the hole energy.  Some examples are shown in
Fig. \ref{fig1} for three values of $q$.  For each $q$ we plot the
integration trajectories in the $(h_x,h_z)$-plane for several values
of $\omega$. The semicircles indicate the moment distribution for
$k_F=250$ MeV.  The nuclear response function, $R_K(q,\omega)$, therefore
correspond to the sum (the integral) of the single-nucleon responses
along one path.  The minimum momentum $h_0$, and therefore the minimum
energy $\epsilon_0$, correspond to the intersection of each curve with
the $h_z$ axis. The curves for different values of $\omega$ do not
intersect. The case $h_0=0$ only occurs for a certain
value of $\omega$, which is precisely the position of the
quasielastic peak; this corresponds also to $\epsilon_0=1$ (or
$\psi^*=0$ for the scaling variable, see below). 
For very large or very small $\omega$-values, the curves lie
in the region where the momentum distribution is zero, and therefore
the corresponding response function is also zero.

Now we define a mean value of the single-nucleon responses 
by averaging with  the
energy distribution $n(\epsilon)$
\begin{eqnarray}\label{def_sn}
\overline{w}_K(q,\omega)
 = \frac{\int^{\infty}_{\epsilon_0} d \epsilon \,  n (\epsilon)
w_K(\epsilon,q,\omega)}{\int_{\epsilon_0}^{\infty} d\epsilon \, n (\epsilon)}.
\end{eqnarray}
This corresponds to the average of the single-nucleon response
$w_K(\epsilon,q,\omega)$ over one of the paths in Fig. \ref{fig1}.
Using these averaged single-nucleon responses
we can rewrite Eq. (\ref{respuesta}) in the form
 \begin{eqnarray}
R_K^{QE}(q,\omega)
&=& 
 \frac{V}{(2 \pi)^3}
 \frac{2\pi m_N^{*3}}{q}  2\overline{w}_K(q,\omega)
\int_{\epsilon_0}^{\infty}d\epsilon\, n(\epsilon).
\label{respuesta2}
\end{eqnarray}
This last integral depends on the
variable $\epsilon_0$, which in turn depends on $(q, \omega)$. 

\subsection{Scaling}
%-------------------------------------------
In the super-scaling approach
 the $\psi^*$-scaling variable is used instead of the minimum energy of
the nucleon, $\epsilon_0$. This energy is transformed by a change of
variable into the scaling variable, $\psi^*$, defined as
\begin{equation}
\psi^* = \sqrt{\frac{\epsilon_0-1}{\epsilon_F-1}} {\rm sgn} (\lambda-\tau),
\end{equation}
where $\psi^*$ is negative (positive) for $\lambda<\tau$ ($\lambda>\tau$).

The superscaling function is defined as
\begin{equation} \label{scaling}
\frac{4}{3}(\epsilon_F-1)
 f^*(\psi^*)=\int_{\epsilon_0}^{\infty} n (\epsilon) d\epsilon,
\end{equation}
where $\epsilon_F-1 \ll 1$ is the kinetic Fermi energy in units of
$m_N^*$.  The definition (\ref{scaling}) is, except for a factor,
similar to that of the $y$-scaling function $f(y)$ \cite{Sar93,Wes75},
where the scaling variable $y$ was the minimum moment of the initial
nucleon.
 
In RFG and nuclear matter with RMF Eq. (\ref{scaling}) is easily evaluated
(remember that the RFG is recovered as the particular case $M^*=1$) as
\begin{equation} 
\int_{\epsilon_0}^{\infty} \theta(\epsilon_F-\epsilon) d\epsilon
= \theta(\epsilon_F-\epsilon_0) (\epsilon_F-\epsilon_0)
= (\epsilon_F-1)(1-\psi^*{}^2)\theta(1-\psi^*{}^2).
\end{equation}
Therefore the scaling function of nuclear matter is
\begin{equation} \label{scalingRFG}
 f^*(\psi^*)=\frac{3}{4}(1-\psi^*{}^2)\theta(1-\psi^*{}^2).
\end{equation}
Note that  the scaling function of nuclear matter is zero for
$\epsilon_0 > \epsilon_F$, and this is equivalent to $|\psi^*| > 1$.
This is a consequence of the maximum momentum $k_F$ for the nucleons
in nuclear matter, which implies that $\epsilon_0< \epsilon_F$.

Using $V/(2\pi)^3= N/(\frac83 \pi k_F^3)$ for nuclear matter
we can write the response
functions (\ref{respuesta2}) as
\begin{equation}
R^{QE}_K(q,\omega) = 
\frac{\epsilon_F-1}{m_N^* \eta_F^3 \kappa}
 (Z \overline{w}^p_K(q,\omega)+N \overline{w}^n_K(q,\omega))
f^*(\psi^*),
\label{susam} 
\end{equation}
where we have added the contribution of $Z$ protons and $N$ neutrons
to the response functions, and $\eta_F=k_F/m_N^*$.

\begin{figure}[h]
\includegraphics[width=9cm,bb=110 270 460 770]{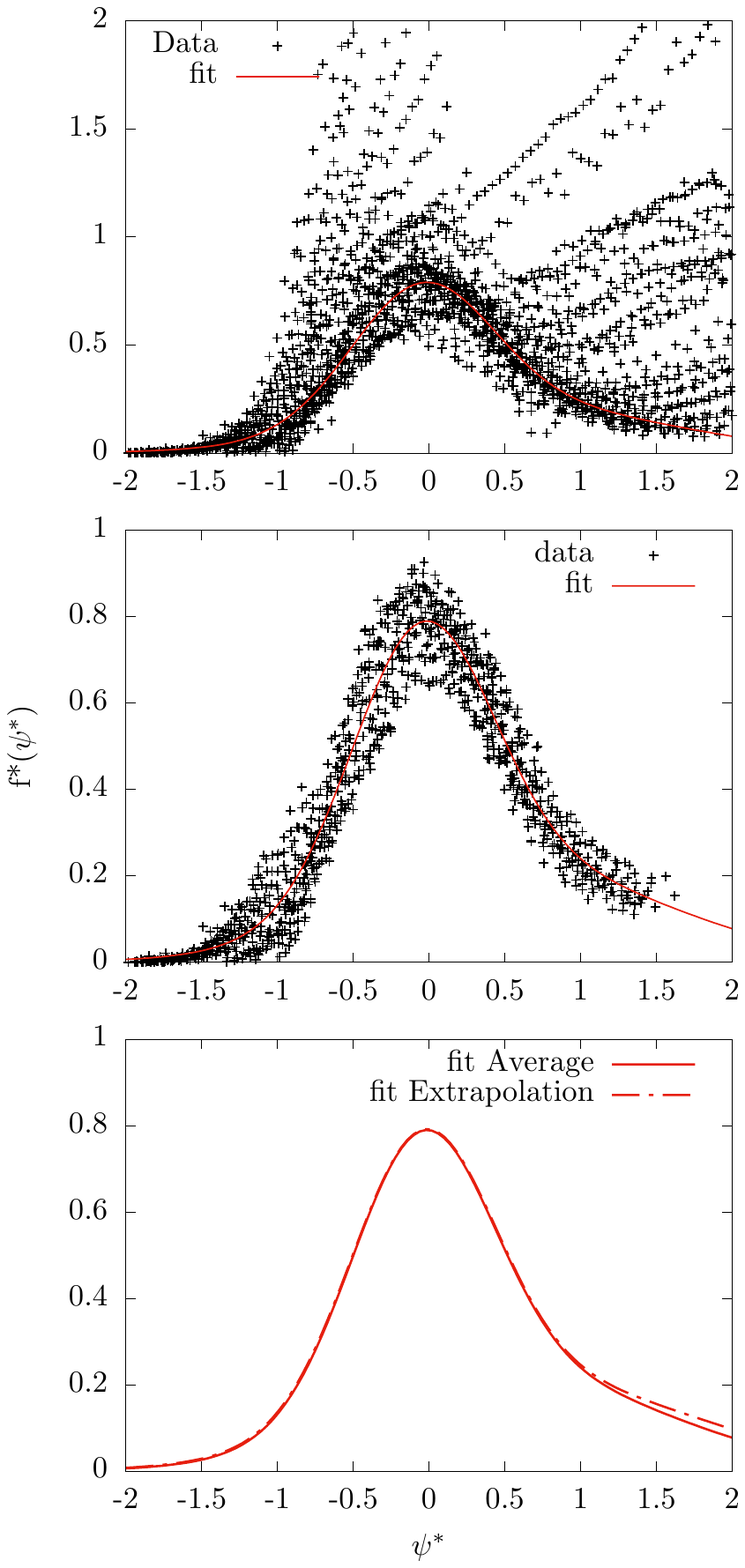}
\caption{
Super scaling analysis with relativistic effective mass
  (SuSAM*) of $^{12}$C data. Top panel: experimental scaling data
$f^*_{exp}$ plotted against $\psi^*$.  Middle panel: data surviving
after cleanup of non-quasielastic sparse points.  The red curve is
Gaussian fit made in this work, $f^*_{QE}(\psi^*)$.  In the bottom panel
we compare the two scaling functions obtained with two different
definitions of the averaged single-nucleon responses: 
using the extrapolated Fermi gas responses and performing the average
with a Fermi distribution defined in Sect 3.
}
  \label{fig0}
\end{figure}

\subsection{SuSAM*}

The SuSAM* approach extends the formula (\ref{susam}) by replacing
$f^*(\psi^*)$ by a phenomenological scaling function obtained from
experimental data of $(e,e')$. 
In a real, finite nucleus the momentum is not limited by $k_F$
(in particular correlated nucleons can greatly exceed the Fermi
momentum). This has the effect that the {\em phenomenological}
superscaling function is not zero for $|\psi^*|>1$,
and therefore takes into account that the nucleons are
not limited by a maximum Fermi momentum.  

Several approaches have been used in the past to obtain a
phenomenological scaling function. In the original SuSA model, based
on the RFG without effective mass, the scaling function was obtained
from the longitudinal response data. In the SuSAv2 model, a scaling
function for the transverse response was also introduced by means of
a RMF theoretical model in finite nuclei. In this paper we will focus
on the SuSAM* model with effective mass where the
phenomenological scaling function is obtained directly from the
quasielastic data of the inclusive cross section.  Different scaling
models with effective mass and without effective mass provide
different scaling functions, but all may reproduce the quasielastic cross
section reasonably well, since they have been fitted to experimental
data.

In the procedure followed in ref \cite{Mar17,Ama18,Rui18}
the inclusive cross section data are divided by the contribution of the
single nucleon.
\begin{equation}
f_{exp}^* =\frac{\displaystyle
 \left(\frac{d\sigma}{d\Omega d\omega}\right)_{exp}}
{\sigma_M ( v_L r_L + v_T  r_T) },
\label{fexp}
\end{equation}
where 
\begin{equation} \label{rsn}
r_K = \frac{\epsilon_F-1}{m_N^* \eta_F^3 \kappa} 
(Z \overline{w}^p_K(q,\omega)+N \overline{w}^n_K(q,\omega)).
\end{equation}
In Fig. 2 these experimental data, $f^*_{exp}$, are plotted against
$\psi^*$ in the interval $-2<\psi^*<2$, which we call {\em the quasielastic 
scaling region}
in this work. 
It is observed that about half of them roughly collapse
forming a thin band around the quasielastic peak. 
 This band constitutes the set of selected
data that can be considered QE and we reject the rest, which mainly
contribute to inelastic processes.  The selected quasielastic data are
well parameterized with a sum of two Gaussians, thus obtaining the
phenomenological quasielastic function $f^*_{QE}$, shown also in Fig. 2.

The SuSAM* model was extended in refs. \cite{Mar21,Mar21b}, by
subtracting the theoretical contribution of the meson-exchange
currents (MEC) in the 2p2h channel from the experimental data before
dividing by the single nucleon, that is 
\begin{equation}
f_{exp}^* =
\frac{\displaystyle \left(\frac{d\sigma}{d\Omega d\omega}\right)_{exp}
  -\left(\frac{d\sigma}{d\Omega d\omega}\right)_{MEC}}{\sigma_M ( v_L r_L + v_T
  r_T) }.
\end{equation}

The resulting SuSAM*+MEC model
provided a somewhat smaller scaling function. However in this work we
use the scaling function (\ref{fexp}) 
without subtraction of MEC, since our focus
will be on the average single nucleon responses.  Both models give
similar results for the quasielastic cross section and we do not want
to complicate the calculation by introducing the 2p2h contribution,
that is not relevant for our further discussion.

%--------------------------------------------
\section{Averaged single-nucleon response functions}
%--------------------------------------------

One of the most confusing aspects in the
superscaling formalism is the definition and meaning of the averaged
single-nucleon response functions for $|\psi^*| > 1$ or, equivalently,
$\epsilon_0> \epsilon_F$, i.e., outside the allowed $\omega$-range of
the Fermi gas. One of the goals of this paper is to shed light on this
matter. Traditionally an extrapolation of the Fermi gas formula has
often been used. In this section we expose the intrinsic theoretical
problems of the Fermi gas extrapolation, and  propose
an alternative definition that is more satisfactory from the
theoretical point of view.

\subsection{RFG extrapolation}

In the traditional superscaling 
approach, first the averaged single-nucleon responses 
$\overline{w}_K$ are calculated 
for $\epsilon_0<\epsilon_F$ (or $|\psi^*|<1$) using the Fermi gas
momentum distribution, 
\begin{eqnarray}
\overline{w}_K(q,\omega)
&=&
\frac{\int_{\epsilon_0}^{\infty}w_K(\epsilon,q,\omega)
      \theta(\epsilon_F-\epsilon)d\epsilon }
{ \int_{\epsilon_0}^{\infty}
\theta(\epsilon_F-\epsilon)d\epsilon} 
\nonumber\\
&=&
\frac{ \theta(\epsilon_F -\epsilon_0) 
\int_{\epsilon_0}^{\epsilon_F}
w_K(\epsilon,q,\omega)}
{\theta(\epsilon_F-\epsilon_0) \int_{\epsilon_0}^{\epsilon_F}d\epsilon}. 
\end{eqnarray}
Note that this expression is only defined for $\epsilon_0 < \epsilon_F$,
in which case the step functions cancel and  we obtain
\begin{equation} \label{wkextrapolated}
\overline{w}_K(q,\omega)=
\frac{1}{\epsilon_F-\epsilon_0}
\int_{\epsilon_0}^{\epsilon_F}
w_K(\epsilon,q,\omega)
d\epsilon
,\kern 1cm
(\epsilon_0 < \epsilon_F).
\end{equation}
The function $w_K(\epsilon,q,\omega)$ inside the integral is well
defined and positive only if $\epsilon>\epsilon_0$,
because it corresponds to the response of a single nucleon with energy
$\epsilon$, that absorbs momentum $q$ and energy $\omega$. 
In the traditional SuSA and SuSAM*
approaches the function (\ref{wkextrapolated}) is extended
analytically for $\epsilon_0> \epsilon_F$ in the obvious way.  This is
called in this work the {\em extrapolated} single nucleon response
function, and it can be written equivalently in the way
\begin{equation}
\overline{w}_K(q,\omega)=
\frac{1}{\epsilon_0-\epsilon_F}
\int_{\epsilon_F}^{\epsilon_0}
w_K(\epsilon,q,\omega)
d\epsilon.
\end{equation}
From this expression it is clear that,for $\epsilon_0 > \epsilon_F$,
the function $w_K(\epsilon,q,\omega)$ inside the integral must be
evaluated for $\epsilon<\epsilon_0$.  But this is not possible for a
nucleon on-shell that absorbs $(q,\omega)$, because its minimum energy
is $\epsilon_0$.  Therefore it is not guaranteed that the function
$w_K(\epsilon,q,\omega)$ inside the integral is positive if is
evaluated for $\epsilon<\epsilon_0$.
 This is a fundamental problem of the single
nucleon extrapolation. Next we will study some particular cases where
the extrapolated responses are 
explicitly negative for $\epsilon_0>\epsilon_F$, that
is, for $|\psi^*|>1$.

\subsection{Longitudinal single-nucleon response}
%-------------------------------------------------

We  use the analytical formulas
of the single nucleon responses from Appendix \ref{appendixA}.
\begin{equation} \label{wl}
w_L= 
\frac{(G_M^*)^2}{1+\tau}
[\tau(\epsilon+\lambda)^2-(1+\tau)\kappa^2]
+\frac{(G_E^*)^2}{1+\tau}
(\epsilon+\lambda)^2.
\end{equation}
To better understand the kinematic dependence of this response function 
it is convenient to express it in terms of the minimal nucleon 
energy $\epsilon_0$ using
\begin{equation}\label{e0lambda}
\epsilon_0+\lambda = \kappa\sqrt{\frac{1+\tau}{\tau}}
\Longrightarrow
\kappa^2(1+\tau)= \tau(\epsilon_0+\lambda)^2.
\end{equation}
in the regime without Pauli blocking. Then Eq. (\ref{wl}) becomes
\begin{equation}
w_L= 
\frac{(G_M^*)^2\tau}{1+\tau}
[(\epsilon+\lambda)^2-(\epsilon_0+\lambda)^2]
+\frac{(G_E^*)^2}{1+\tau}
(\epsilon+\lambda)^2.
\end{equation}
In this equation it is evident that the electric term is always
positive.  However the magnetic term is positive only for $\epsilon >
\epsilon_0$.  For this reason, if $w_L$ is calculated using the Fermi
gas momentum distribution and then extrapolated to values
$\epsilon_0>\epsilon_F$ (or $\psi^*>1$), the magnetic term becomes
negative. This does not make physical sense because the longitudinal
response must be positive, by definition, regardless of the value of
the form factors. In fact if we artificially turn off the electric
contribution, a negative averaged response $\overline{w_L}$ is
obtained for $\epsilon_0> \epsilon_F$.  Let suppose for
simplicity that $G_E^*= 0$.  Then the extrapolated
single-nucleon longitudinal response would be
\begin{equation}\label{wlM}
\overline{w}_L=
\frac{(G_M^*)^2}{\epsilon_0-\epsilon_F}
\frac{\tau}{1+\tau}
\int_{\epsilon_F}^{\epsilon_0}
[(\epsilon+\lambda)^2-(\epsilon_0+\lambda)^2]d\epsilon,
\end{equation}
that 
is negative for $\epsilon_0 > \epsilon_F$.

\subsection{Transverse single-nucleon response}

We find a similar situation in the case 
of the transverse response  from Eq. (\ref{wtfinal}) in the Appendix
 \ref{appendixA}
\begin{equation}
w_T= 
2\tau (G_M^*)^2
+\frac{(G_E^*)^2+\tau (G_M^*)^2}{1+\tau}
\frac{\tau}{\kappa^2}
\left[(\epsilon+\lambda)^2-\kappa^2\frac{1+\tau}{\tau}\right].
\end{equation}
Again we can rewrite this response as a function of the minimum 
nucleon energy, $\epsilon_0$,
using $\kappa^2(1+\tau)/\tau= (\epsilon_0+\lambda)^2$
\begin{equation}
w_T= 
2\tau (G_M^*)^2
+\left[(G_E^*)^2+\tau (G_M^*)^2\right]
\left[
\left(\frac{\epsilon+\lambda}{\epsilon_0+\lambda}\right)^2-1
\right].
\end{equation}
Rearranging terms containing $G_E^*$ and $G_M^*$ the single-nucleon transverse 
response becomes finally
\begin{equation} \label{wtexplicit}
w_T= 
(G_E^*)^2
\left[
\left(\frac{\epsilon+\lambda}{\epsilon_0+\lambda}\right)^2-1
\right]
+\tau (G_M^*)^2
\left[
\left(\frac{\epsilon+\lambda}{\epsilon_0+\lambda}\right)^2+1
\right].
\end{equation}
Written in this way, it is evident that the magnetic contribution of
$w_T$ is always positive. While the electrical term is positive only for
$\epsilon > \epsilon_0$.
The situation is similar to what we found with the longitudinal
response, but in the transverse response it is the electrical term
that becomes negative in the extrapolation to $\epsilon_0>\epsilon_F$. 
We now turn off the
magnetic contribution and suppose that $G_M^*=0$.  Then
the averaged T response in RFG would be, with analogy to
Eq. (\ref{wlM})
\begin{equation} \label{wtE}
\overline{w}_T=
\frac{(G_E^*)^2}{\epsilon_0-\epsilon_F}
\int_{\epsilon_F}^{\epsilon_0}
\left[
\left(\frac{\epsilon+\lambda}{\epsilon_0+\lambda}\right)^2-1
\right]
d\epsilon.
\end{equation}
From this expression it is clear that the extrapolated
$\overline{w}_T$ is negative for $\epsilon_0 > \epsilon_F$ because the
function inside the integral is negative, which is not physically acceptable:
the transverse response should be
positive by definition regardless of the form
factors values. In other words, the electrical contribution to the transverse
response, although samll, cannot be negative.

\subsection{Alternative to the extrapolated single-nucleon responses}

In this work we propose an alternative definition of the averaged
single-nucleon responses that solves the extrapolation problem in the
superscaling model.  As we have seen, the problem is a consequence of
the fact that in the Fermi gas there is a maximum momentum for the
nucleons. If this momentum is exceeded by extrapolation, i.e.
$\epsilon_0 > \epsilon_F$, mathematically this is equivalent to
assuming nucleons with energy less than $\epsilon_0$, which is
impossible in the Fermi gas because nucleons are on-shell.  Hence
results without physical sense, such as negative responses, are
obtained if the extrapolated formula is applied.

The proposed solution involves using equation (13) for the averaged
single-nucleon responses, but introducing a momentum distribution
without a maximum momentum, and that at the same time does not differ
much from the Fermi gas distribution, for $h<k_F$. An appropriate
function is a distribution of Fermi type
\begin{equation}\label{distribucion}
n(h)= \frac{a}{1+e^{(h-k_F)/b}}.
\end{equation}
Where $b$ is a smearing parameter for the Fermi surface, which is no
longer restricted to a sphere as in figure 1. Then the integrals by
averaging in Eq. (13) extend to infinity and therefore there is no
longer an upper limit for $\epsilon_0$, which can take any value up to
infinity. The single-nucleon responses 
of the integrand always are evaluated for $\epsilon
> \epsilon_0$ and they are therefore positive definite (see
eqs. (28,32)).

\begin{figure}[t]
\includegraphics[width=8cm,bb=110 270 460 770]{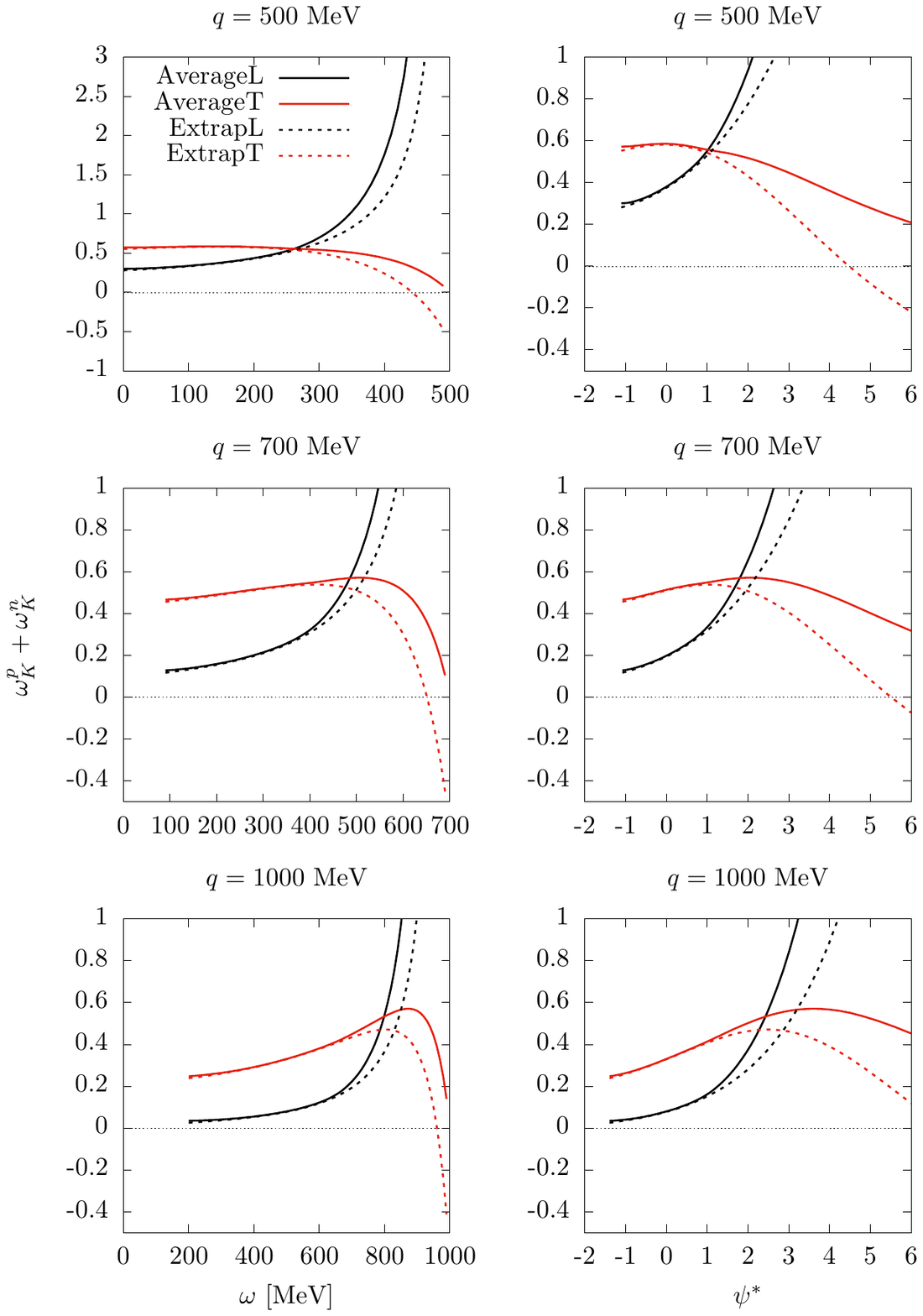}
  \caption{Averaged and extrapolated longitudinal and transverse
    response functions for proton plus neutron, as a function of
    $\omega$ and of the scaling variable $\psi^*$, for three values of
    the momentum transfer.}
  \label{fig3}
\end{figure}

Besides, for $\epsilon < \epsilon_F$, the momentum distribution is
similar to the Fermi Gas distribution, $\theta(k_F-h)$, and then it is
expected that the averaged single-nucleon be similar to that of the
RFG (see Fig.1).  Now, for $\epsilon > \epsilon_F$ the integration
(13) extends in momentum space along one of the paths outside the
Fermi sphere of Fig. 1. Then the average has the physical sense of
coming from regions above the Fermi sphere, that is to say, from the
high momentum zone that the Fermi gas cannot describe. This is in
accordance with the meaning attached to the experimental scaling
function for $|\psi^*|>1$, which comes mainly from high momentum
nucleons.

%-------------------
\section{Results}
%--------------------

In this section we present results for the averaged nucleon responses
and for the total nuclear responses in the SuSAM* model. The
calculations are made for electron scattering off the nucleus $^{12}$C
with Fermi momentum $k_F=225$ MeV/c and effective mass
$m_N^*=0.8m_N$. These values were fitted to the quasielastic data of
$f^*_{exp}$ to obtain the best possible scaling \cite{Mar17,Ama18}.
We evaluate the validity of the scaling model when using the Fermi gas
extrapolation for the nucleon response function. Specifically,
the results obtained by averaging the single-nucleon response
function over a smeared Fermi momentum distribution,
Eq. (\ref{distribucion}) are compared with the extrapolated response function
obtained from the Fermi gas model.

In Fig. \ref{fig3} we compare the averaged nucleon responses with the
extrapolated ones. The sum of proton plus neutron is shown. The
averaged responses have been calculated with a Fermi distribution
using a smearing parameter $b=50$ MeV/c.  The responses do not depend
much on the precise value of this parameter for small variations. We
see that the averaged responses are practically the same as the
extrapolated responses of the Fermi gas 
in the quasielastic scaling region, $-2<\psi^*<2$,.
 But both results start to diverge for large $\omega$ or
$\psi^*>2$. The extrapolated transverse response
becomes negative for $\psi^* > 4$, 5 and 7, for $q=500$, 700, and
1000 MeV/c, respectively, very close to the photon line. 
This is easily explained because in Eq. (32)
the magnetic term is multiplied by $\tau$.  Therefore the $w_T$
response is dominated by the electric term for $\tau\rightarrow 0$,
that is, for large $\omega$, and in Eq. (33) we have seen that this
term is negative when extrapolated to $\epsilon_0 > \epsilon_F$.

\begin{figure}[t]
\centering
\includegraphics[width=7cm,bb=110 270 460 770]{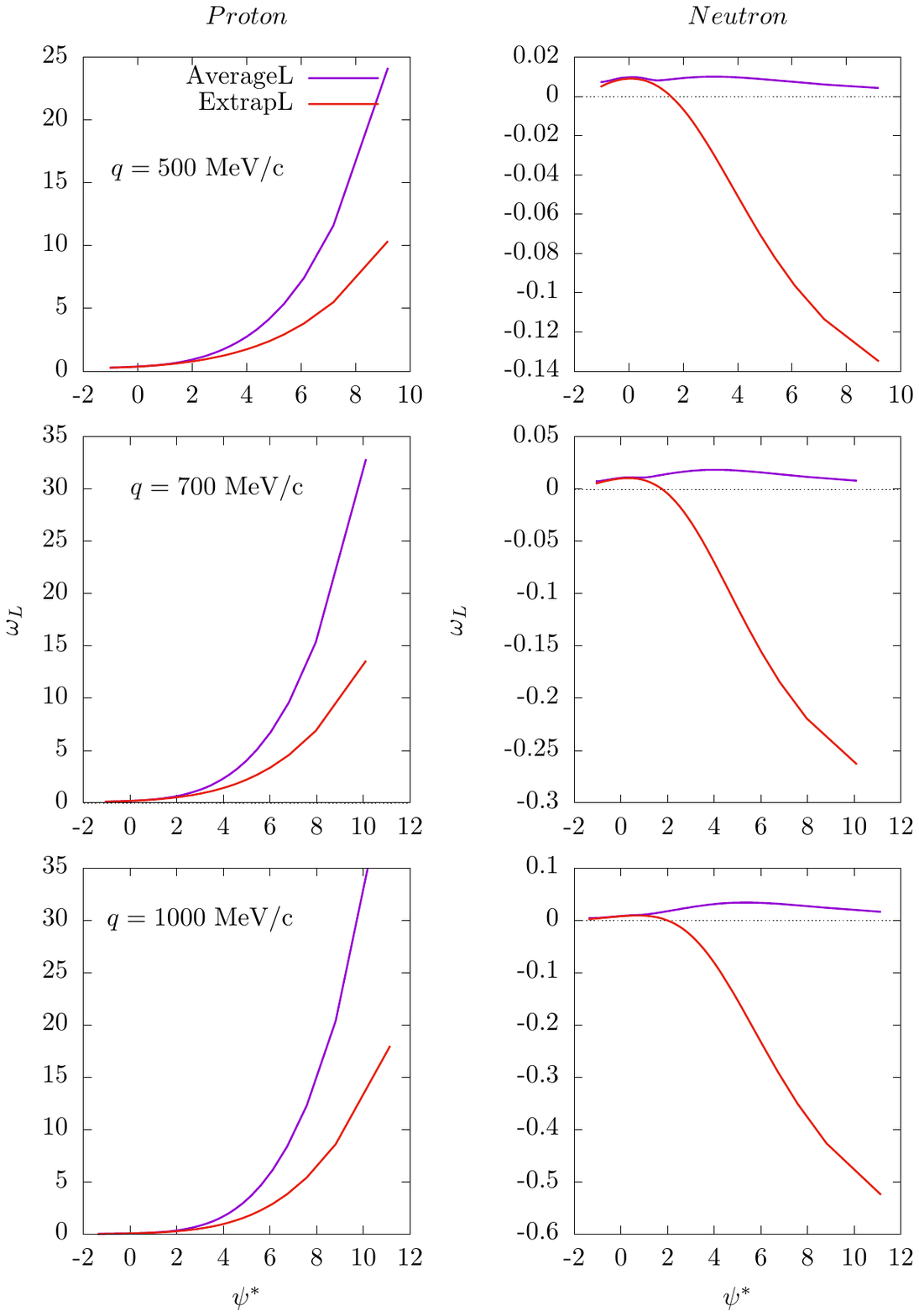}
\kern 5mm
\includegraphics[width=7cm,bb=110 270 460 770]{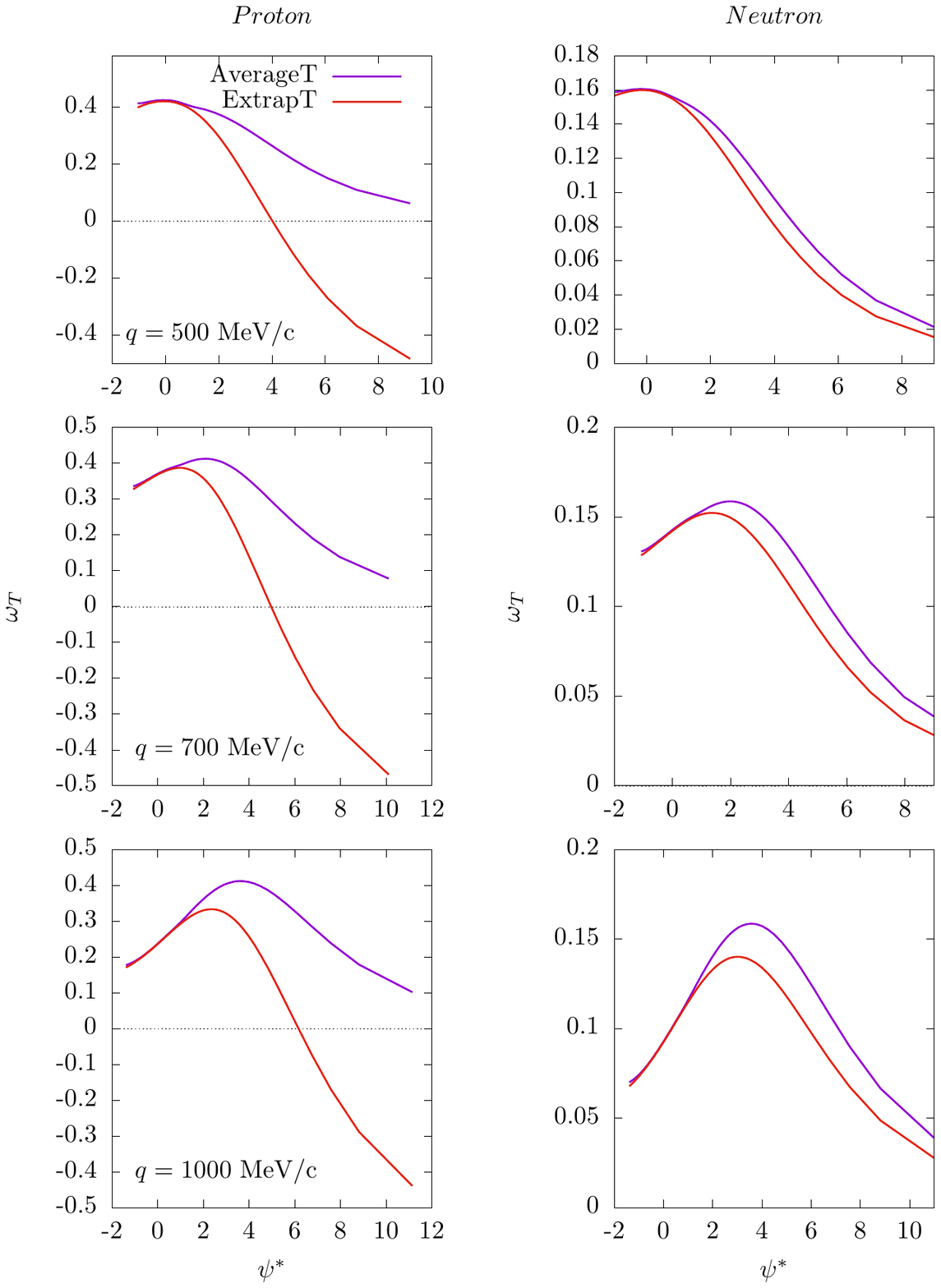}
  \caption{Averaged and extrapolated longitudinal and transverse 
response functions for
    protons and neutrons, as a function of the scaling
    variable and for three values of the momentum transfer.}
  \label{fig4}
\end{figure}

More details can be seen in Fig.  \ref{fig4} where we show the
averaged and extrapolated response functions separated for protons and
neutrons, as a function of the scaling variable.  The extrapolated and
averaged responses start to differ in the region $\psi^*>2$ and the
discrepancy increases with $\psi^*$.  The extrapolated longitudinal
response of neutrons is negative for $\psi^* >2$. This agrees with
what was seen analytically in the previous section, because the
extrapolation of the longitudinal magnetic response is negative and
the electric form factor of the neutron is negligible.  This does not
affect the results of the SuSAM* model in the scaling region
 because the longitudinal
response of the neutron is much smaller than that of the proton.

In fig. \ref{fig4} we also can see see that the averaged proton
transverse response is very similar to the extrapolation in the
scaling region and differ for $\psi^*>2$.  
They also start to differ in the
$\psi^*$-negative region for $\psi^* < -2$.  The extrapolated
transverse response of protons is negative from $\psi^*\sim 4$--6
depending on the value of $q$.  Again this is because the electrical
term of the proton dominates this response for large $\omega$ since
the magnetic term carries a factor $\tau$, which tends to zero for
$\omega \rightarrow q$. In contrast the averaged proton transverse
responses are always positive.

The averaged transverse neutron response shown in Fig. \ref{fig4} is
similar in shape to the Fermi gas extrapolation in the scaling region. 
But again they differ for
$|\psi^*|>2$, where the averaged one is the largest, and the difference
between the two increases with the momentum transfer.

\begin{figure}[t]
\centering
\includegraphics[width=7cm,bb=110 270 460 770]{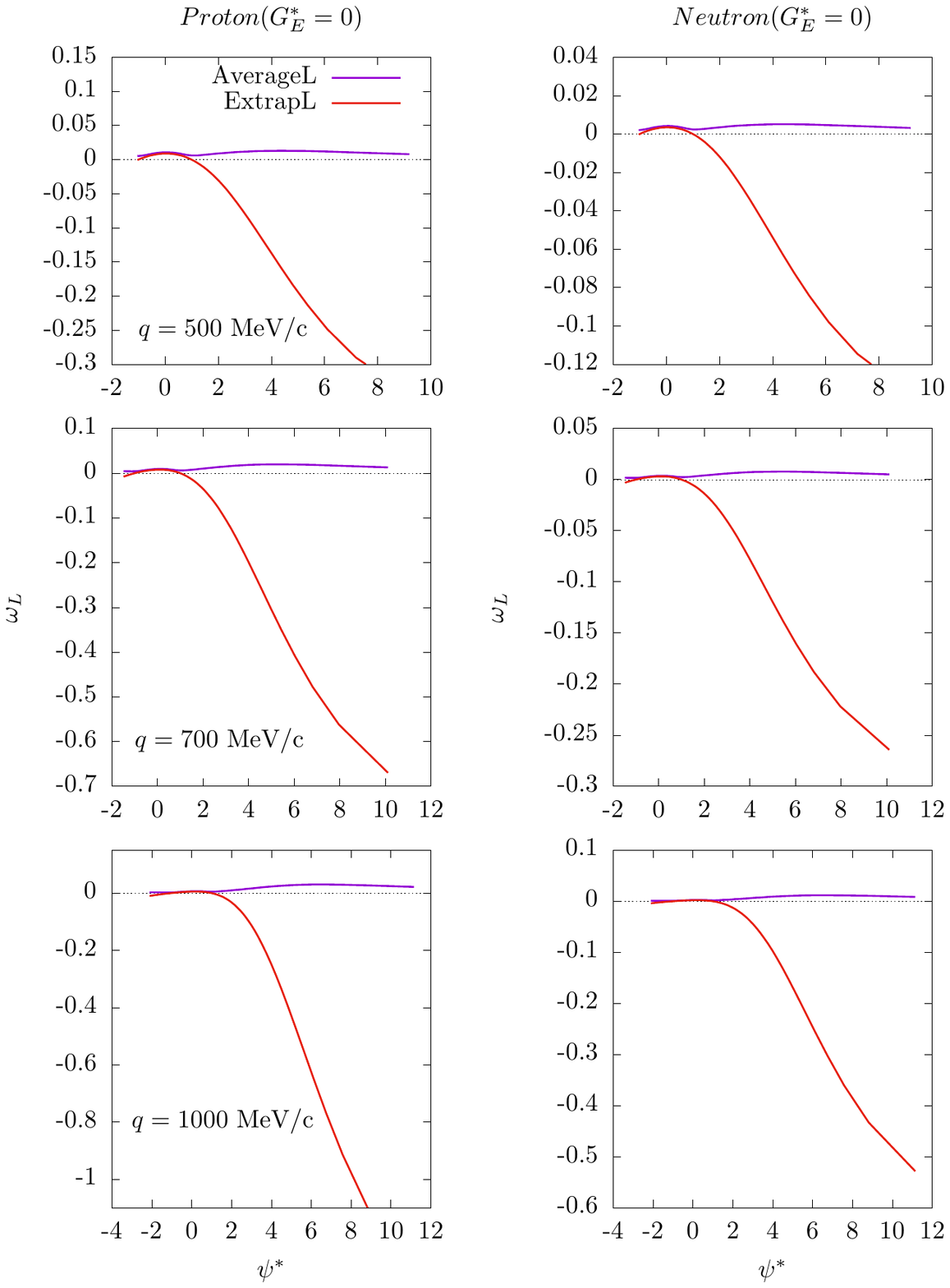}
\kern 5mm
\includegraphics[width=7cm,bb=110 270 460 770]{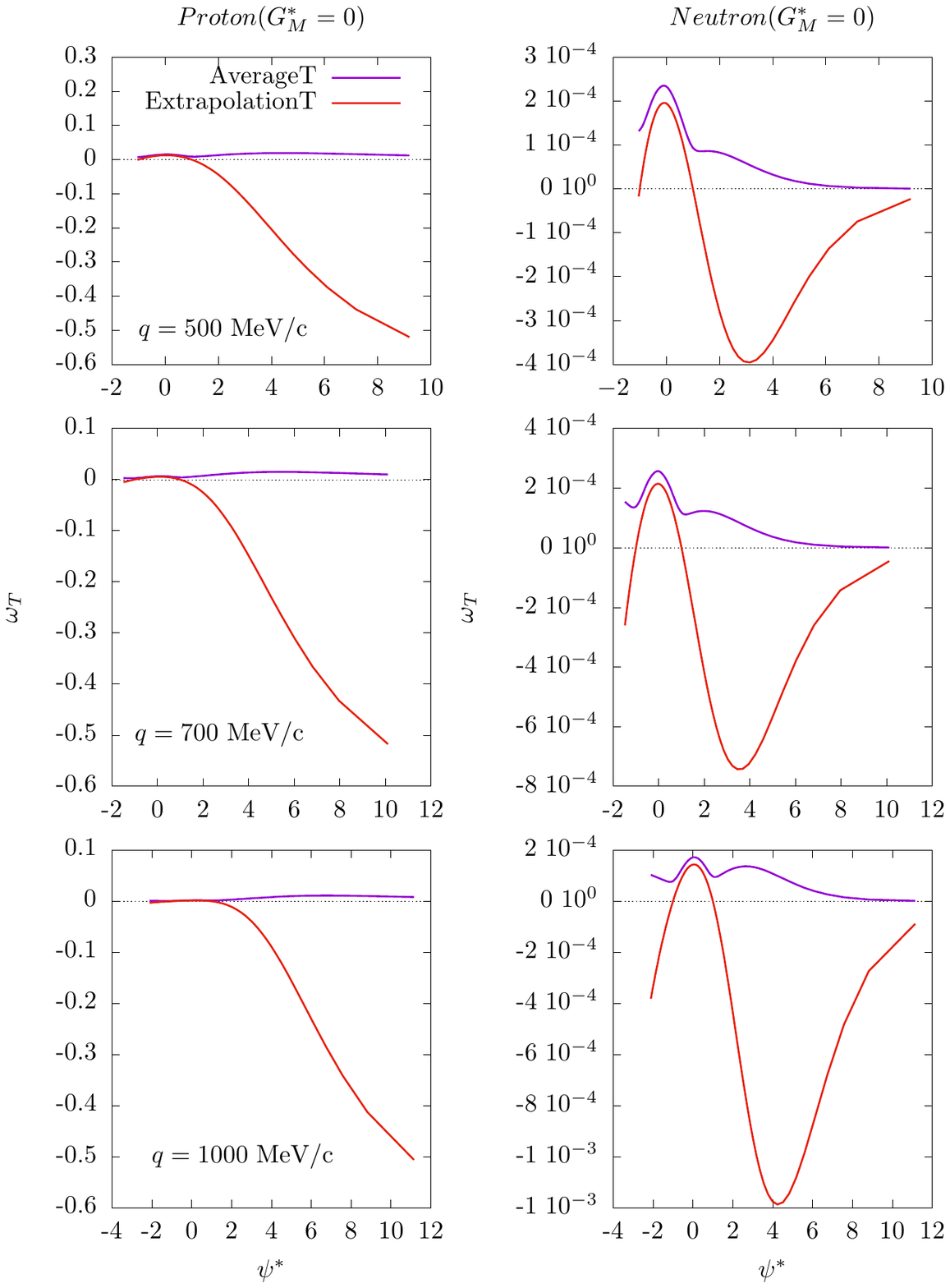}
  \caption{Averaged and extrapolated transverse response functions for
    protons and neutrons, for $G_M^*=0$, as a function of the scaling
    variable and for three values of the momentum transfer.
     Averaged and extrapolated longitudinal response functions for
    protons and neutrons, for $G_E^*=0$, as a function of the scaling
    variable and for three values of the momentum transfer.}
  \label{fig5}
\end{figure}

We have seen in the extrapolation formulas, Eqs. (29,33), that the
magnetic contribution to the longitudinal response and the electrical
contribution to the transverse response become both negative for
$\epsilon_0> \epsilon_F$, This can be explicitly seen in the results
in Fig. \ref{fig5}, where we plot the longitudinal responses computed
for $G^*_E=0$ and the transverse responses computed for $G^*_M=0$,
for protons and neutrons.  In fact, in all cases of Fig. \ref{fig5}
the extrapolated responses are negative for $|\psi^*|>1$. On the
contrary the averaged responses are always positive.

\begin{figure}
\centering
\includegraphics[width=7cm,bb=110 435 460 785]{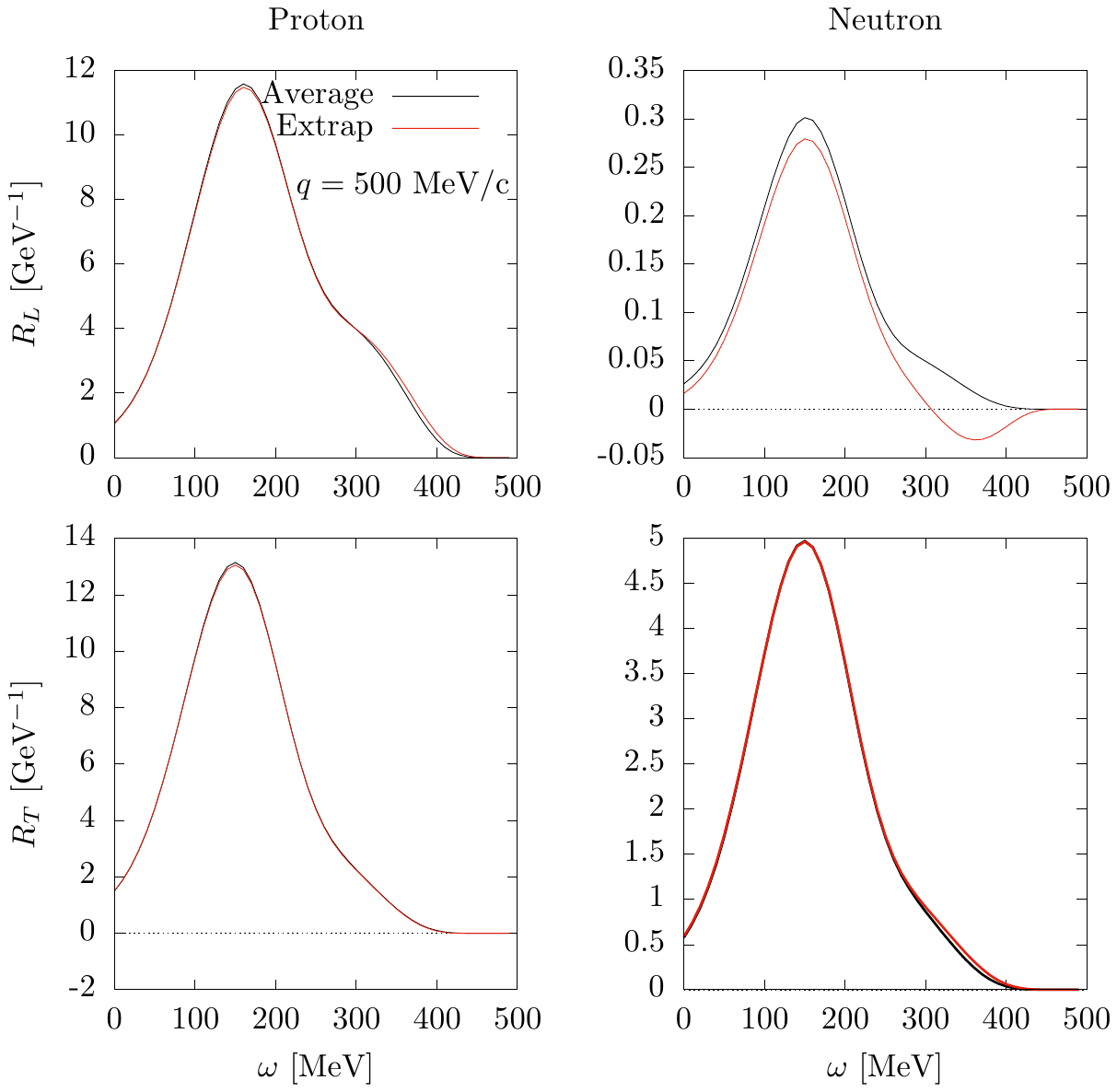}
\kern 4mm
\includegraphics[width=7cm,bb=110 435 460 785]{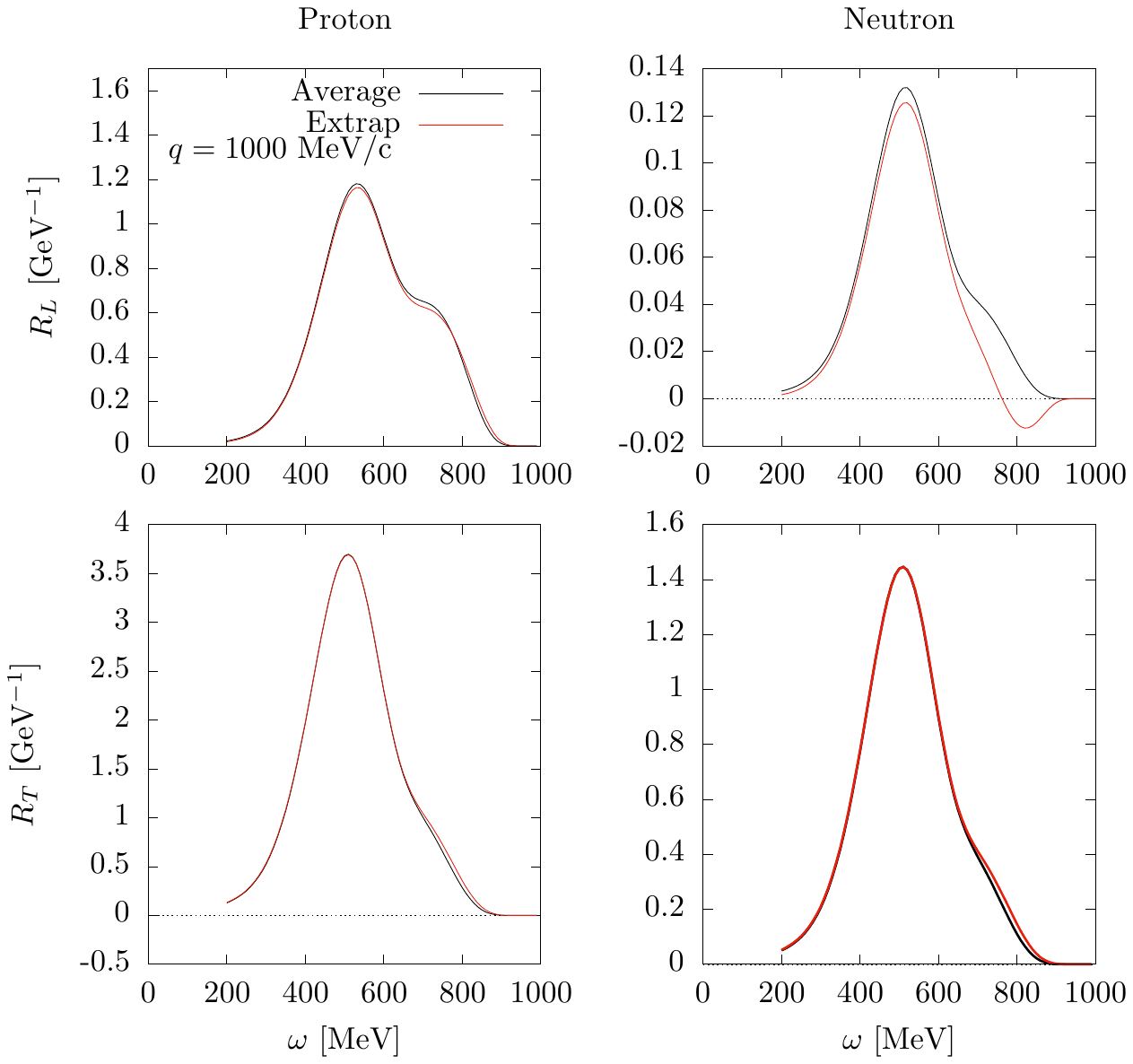}
  \caption{Longitudinal and transverse response functions separated 
for protons
    and neutrons in the SuSAM* model using the averaged and
    extrapolated single nucleon responses for $q=500$ (two left panels) 
and 1000 MeV/c (two right panels).}
  \label{fig6}
\end{figure}

In fig \ref{fig6} we use the superscaling model to investigate the
nuclear responses under various inputs for the single-nucleon.  The
nuclear response is computed from the product of the averaged
nucleon-responses and a phenomenological scaling function obtained
from the data, using Eq. (19).

The results in Fig. \ref{fig6} demonstrate that there are no significant
differences in the separate responses of protons and neutrons when
computed with the averaged single-nucleon compared to the
extrapolation. The only difference is seen in the longitudinal neutron
response for high $\omega$, which becomes negative in the extrapolated
model.  However this is not relevant for the total nuclear response,
as the neutron contribution is negligible in the longitudinal response
as compared to the proton one.

\begin{figure}
\includegraphics[width=8cm,bb=110 270 460 770]{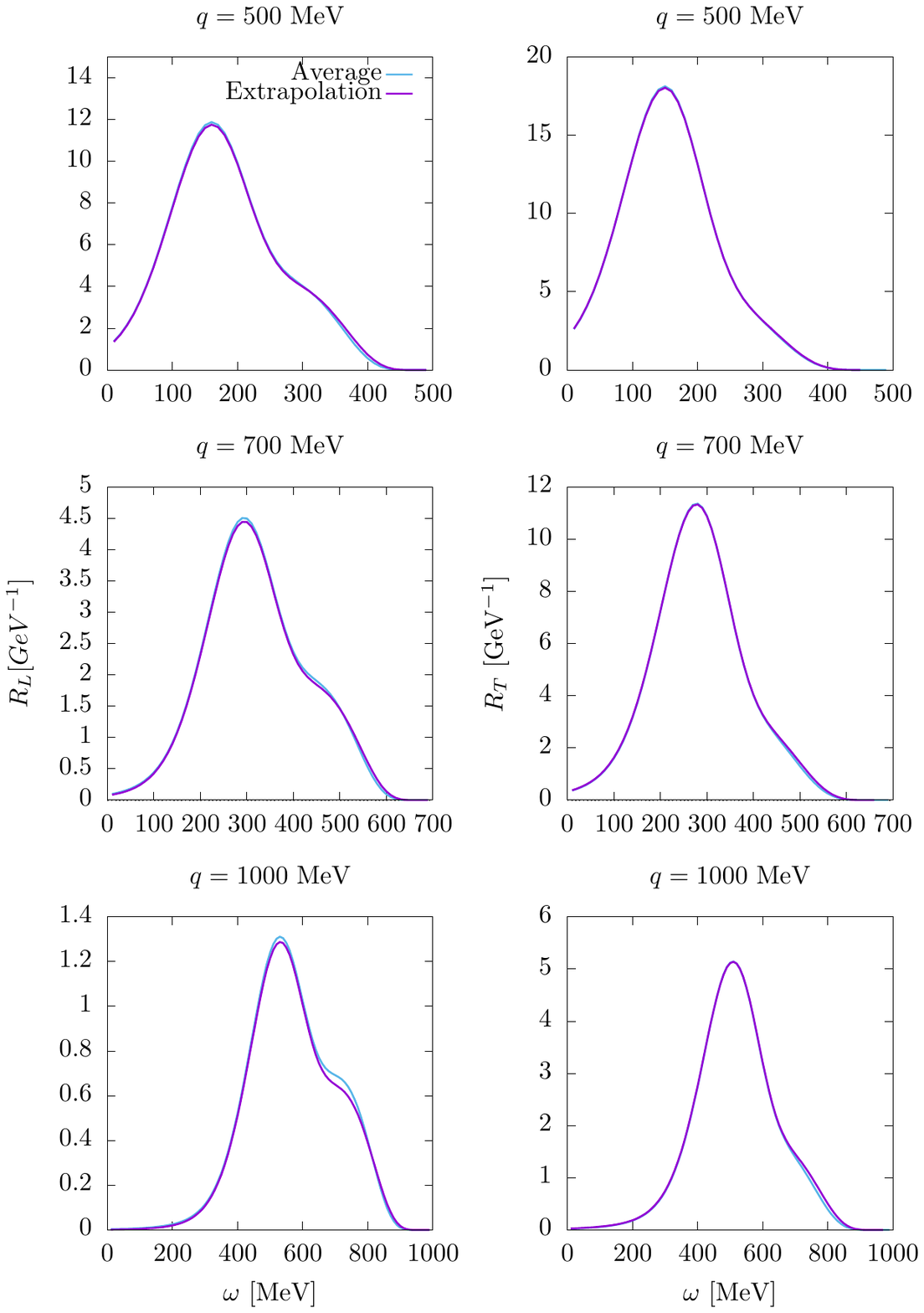}
  \caption{Longitudinal and transverse response functions in the SuSAM* 
model using the averaged and extrapolated single nucleon responses}
  \label{fig7}
\end{figure}

This is verified in the results of Fig. \ref{fig7} for the total
responses. Both the averaged and the extrapolated single-nucleon
responses give essentially the same result.  The results obtained 
have two important implications. Firstly, they provide
support for the validity of using the single-nucleon response
extrapolated from the Fermi gas, as this approach yields the same
results as using a response averaged with a nuclear momentum
distribution that does not have a maximum momentum. Secondly, they
justify the use of the averaged response as a means of avoiding the
potential issues that we have identified with the extrapolation
method.

Finally we have conducted a new scaling analysis of the $^{12}C$ data
using the single-nucleon response averaged with the Fermi
distribution. The results, as shown in Figure 2, demonstrate that the
scaling function obtained using this approach is virtually
indistinguishable from the one obtained through extrapolation. These
findings highlight the robustness of the scaling approach and suggest
that using the averaged response may be a viable alternative to
extrapolation in certain cases.  Furthermore, in Figures 8 and 9, we
compare the cross-section of $^{12}$C using the SuSAM* model and the
RMF model of nuclear matter for a selected set of
kinematics. The SuSAM* model still proves to be an excellent method
to parameterize the quasielastic cross-section through a single
scaling function.

%%%%%%%%%%%%%%%%%%%%%%%%%%%%%%%%%%%%%%%%%%
\section{Discussion}

The findings of the results section demonstrate the robustness and
versatility of the superscaling models with respect to the choice of
the averaged single-responses, and its potential applications in a
variety of situations in electron and neutrino scattering.  The
updated single-nucleon responses provide a well-defined
theoretical basis for the scaling function that is compatible with the
traditional extrapolation in the scaling region.  This reinforces the
universality of the scaling function because it is independent of the
way in which the average response of the nucleon is defined. This
means that the scaling function can be used to describe the
electromagnetic response of nucleons in different types of nuclei,
regardless of their size or composition.

\begin{figure}
\includegraphics[width=12cm,bb=40 275 530 775]{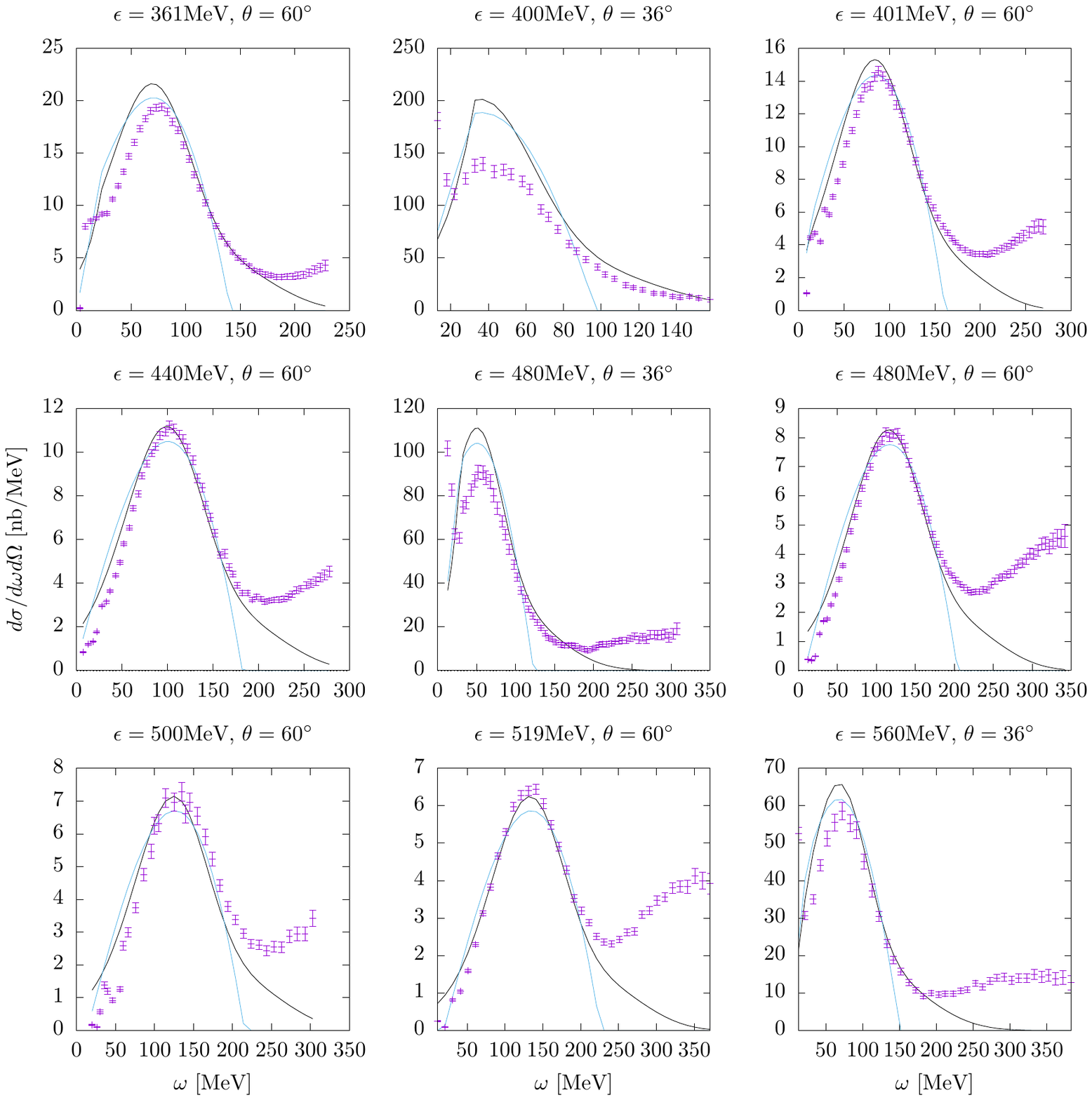}
  \caption{Quasielastic $(e,e')$ cross section of $^{12}C$ as a
    function of $\omega$ for several values of the electron energy,
    $\epsilon$, and scattering angles $\theta$, computed with the
    present SuSAM* model (black lines) compared to the RFG with
    effective mass (blue lines). Experimental data are from
    refs. \cite{archive,archive2}}
  \label{fig12}
\end{figure}

The averaged single-nucleon model has promising applications in other
situations outside the scaling region for high-energy transfer. For
instance in two-particle emission reactions, two-particle two-hole
(2p2h) excitation can be produced by the one-body current due to
nuclear short-range correlations.  The electromagnetic interaction
with a nucleon belonging to a correlated pair can result in the
emission of both nucleons because the correlated nucleons acquire
high-momentum components that allow the overlap of the wave function
with states above the Fermi momentum.  A simple model of emission of
two correlated nucleons has been proposed in ref. \cite{Mar22} to
explain phenomenologically the tail of the scaling function at high
energies.  The probability of emission of a proton-neutron pair is
approximated by a factorized model, similar to the scaling approach.
One factor is the sum of the averaged proton and neutron responses
considered in this work.  The other factor is the probability of
emitting two particles while conserving energy and momentum, assumed
to be proportional to the phase space of two particles in the Fermi
gas.  The total response is assumed to be the product of these two
factors with an additional correlation factor $c_{pn}(q)$ that
accounts for the average probability of the high momentum
proton-neutron correlated pair.  The factor $c_{pn}(q)$ is obtained
phenomenologically by fitting the tail of the scaling function.  In
such 2p2h correlation model the contribution of the single nucleon for
high $\omega$ outside of the scaling region $\psi^*>2$ plays an important role, and
the extrapolation of the Fermi gas single-nucleon model is not
appropriate.

\begin{figure}[t]
\includegraphics[width=12cm,bb=40 275 530 775]{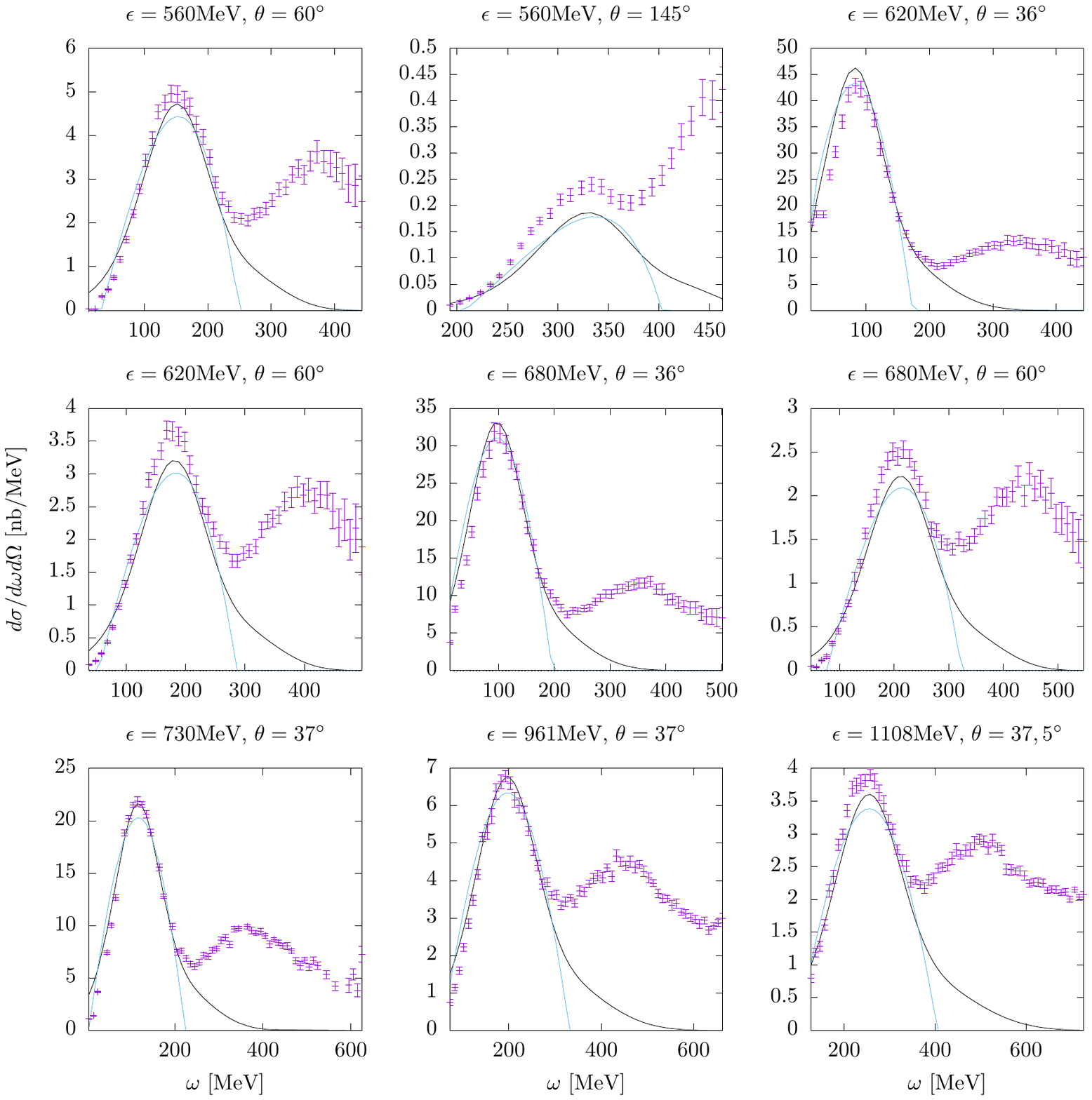}
  \caption{Quasielastic $(e,e')$ cross section of $^{12}C$ as a
    function of $\omega$ for several values of the electron energy,
    $\epsilon$, and scattering angles $\theta$, computed with the
    present SuSAM* model (black lines) compared to the RFG with effective
    mass (blue lines). Experimental data are from refs. \cite{archive,archive2}}
  \label{fig13}
\end{figure}

Another direct application of this method concerns the calculation of
the contribution of meson-exchange currents (MEC) to the quasielastic
1p1h response in the superscaling model. This calculation was
performed in the RFG for instance in Refs. \cite{Ama01,Ama03} and
involves computing an effective one-body current as the sum of
one-body plus MEC, 1p1h matrix elements.  The traditional scaling
model with extrapolation is not trivial to apply in this case, as the
single-nucleon responses of the MEC must be computed
numerically. However, the averaged single-nucleon responses of the
OB+MEC operator can be directly computed as we have done in this work.

\section{Conclusions}

In this work, we have re-examined the scaling formalism from a
theoretical standpoint, with a particular emphasis on the definition
of the averaged electron-nucleon responses, which are assumed to
factorize in the model.  Within the SuSAM* model, which takes into
account the relativistic mean field through the effective mass of the
nucleon, we have investigated the validity of the traditional approach
of extrapolating to $|\psi^*|>1$ the single-nucleon responses averaged
over the Fermi gas.  A detailed analysis shows that that, for
$|\psi^*|>1$, the extrapolation formulas produce nonphysical negative
results for the responses in some particular cases, which contradict
the physical expectation that the response functions should always be
positive.  Specifically, the magnetic contribution of the longitudinal
response and the electrical contribution of the transverse response
become negative for $|\psi^*|>1$.  This is propagated to the total
responses, resulting in the extrapolated single-nucleon transverse response
becoming negative for very high values of $\omega$. 

Therefore we have proposed a different definition for the averaged
single-nucleon responses with a smeared momentum distribution around
the Fermi surface.  This approach does not suffer from the problems
associated with the extrapolation method, and on the other hand
produces results that are similar to those of the extrapolated SuSAM*
model.  Our proposed approach, which takes into account the
high-momentum nucleons to a certain extent, does not depend
significantly on the fine details of the nuclear density due to the
averaging procedure.  

Despite the theoretical problems with extrapolation, in this work, we
have shown that the extrapolated model produces results similar to the
correctly averaged model within the scaling region $-2< \psi^* < 2$.
In conclusion, we have provided a solid basis for the traditional
superscaling model in the quasielastic peak region.  The new
physically motivated definition of the averaged single-nucleon
responses strengthens the physical interpretation of the superscaling
model for understanding the response of atomic nuclei in electron and
neutrino scattering experiments.  The new method has other
applications as well, for example, it allows for the inclusion of the
MEC effect in the superscaling model. A calculation of this type will
be presented in the future, demonstrating the versatility and
potential of this new approach.

\section{Acknowledgments}

Work supported by:
Grant PID2020-114767GB-I00
funded by MCIN/ AEI/ 10.13039/ 501100011033; FEDER/Junta de
Andalucia-Consejeria de Transformacion Economica, Industria,
Conocimiento y Universidades/A-FQM-390-UGR20; and Junta de
Andalucia (Grant No. FQM-225).

\appendix

\section{Single nucleon responses}
\label{appendixA}

The single-nucleon hadronic tensor 
is computed performing the spin traces 
(\ref{traza}) with the current matrix elements (\ref{corriente}),
and can be written as
\begin{equation}
w^{\mu\nu}= -w_1\left(g^{\mu\nu}-\frac{Q^\mu Q^\nu}{Q^2}\right)+
w_2V^\mu V^\nu,
\label{wmunu}
\end{equation}
where we have defined the four-vector $V^\mu= (H^\mu+Q^\mu/2)/m_N^*$,
and $H^\mu=(E,\nh)$ is the initial nucleon four-momentum with
effective mass $m_N^*$. The four-momentum of the final nucleon is
$P^\mu=H^\mu+Q^\mu$.
The nucleon structure functions are given by
\begin{eqnarray}
w_1(Q^2) &=& \tau (G_M^*)^2 > 0,  \label{w1}\\
w_2(Q^2) &=& \frac{(G_E^*)^2+\tau (G_M^*)^2}{1+\tau} > 0, \label{w2}
\end{eqnarray}
where the electric and magnetic form factors for nucleons with effective mass 
are \cite{Ama15}
\begin{equation}
G_E^*  =  F_1-\tau \frac{m^*_N}{m_N} F_2, \kern 1cm
G_M^*  = F_1+\frac{m_N^*}{m_N} F_2.  \label{GM}
\end{equation}
For the $F_i$ form factors of the nucleon, we use the Galster
parametrizations \cite{Gal71}.  

Note that $w_1$ and $w_2$ are positive and depend only on $Q^2$.
Here we compute the longitudinal and transverse 
components of the hadronic tensor,
$w_L= w^{00}$ and $w_T=w^{11}+w^{22}$ respectively, 
 appearing in inclusive electron scattering.

\paragraph{\bf Longitudinal single-nucleon response}
%-------------------------------------------------
We use the following results for the time components of the basic tensors
and vectors in terms of adimensional variables, $\kappa,\lambda,\tau$
\begin{equation}
g^{00}-\frac{Q^0 Q^0}{Q^2}= -\frac{q^2}{Q^2}= \frac{\kappa^2}{\tau}, 
\end{equation}
\begin{equation}
V^0= \frac{E+\omega/2}{m_N^*}=\epsilon+\lambda.
\end{equation}
Substituting the values of these time components and of the structure
functions in the hadronic tensor (\ref{wmunu}), the longitudinal single-nucleon
response function becomes
\begin{equation}
w_L= 
-\kappa^2(G_M^*)^2
+\frac{(G_E^*)^2+\tau (G_M^*)^2}{1+\tau}(\epsilon+\lambda)^2.
\end{equation}
Rearranging terms containing $G_E^*$ and $G_M^*$ this becomes
\begin{equation} \label{wl}
w_L= 
\frac{(G_M^*)^2}{1+\tau}
[\tau(\epsilon+\lambda)^2-(1+\tau)\kappa^2]
+\frac{(G_E^*)^2}{1+\tau}
(\epsilon+\lambda)^2.
\end{equation}

\paragraph{\bf Transverse single-nucleon response}

In the case of the transverse response  
$g^{ii}=-1$ and $V^i= h_i/m_N^* = \eta_i$, for $i=1,2$, 
where we have defined the 
three-vector $\neta=\nh/m_N^*$. Then
 the T response is  
\begin{equation}\label{wt}
w_T= w^{11}+w^{22}= 2w_1+w_2(\eta_1^2+\eta_2^2).
\end{equation}
Note that
$\eta_1^2+\eta_2^2=\eta^2-\eta_3^2= \epsilon^2-1-\eta_3^2$. 
The value of $\eta_3^2$ is the projection
of the vector $\neta$  over the $\nq$ direction,  
which is determined by energy-momentum conservation. In fact,
using Eq (\ref{angulo}) 
\begin{equation}
\eta_3= \frac{h\cos\theta}{m_N^*}= \frac{E\omega+Q^2}{m_N^* q}
= \frac{\epsilon\lambda-\tau}{\kappa}.
\end{equation}
Then we have
\begin{equation}
\eta_1^2+\eta_2^2= \epsilon^2-1-
\left(\frac{\epsilon\lambda-\tau}{\kappa}\right)^2.
\end{equation}
Expanding the square and  using $\kappa^2-\lambda^2=\tau$, this gives
gives
\begin{equation}
\eta_1^2+\eta_2^2= \frac{\tau}{\kappa^2}
\left[\epsilon^2-\frac{\kappa^2}{\tau}-\tau+2\epsilon\lambda\right]=
\frac{\tau}{\kappa^2}
\left[(\epsilon+\lambda)^2-\kappa^2\frac{1+\tau}{\tau}\right].
\end{equation}
Inserting this result in Eq. (\ref{wt}) and using the values of 
$w_i$ from Eqs. (\ref{w1},\ref{w2}), the transverse response becomes
\begin{equation} \label{wtfinal}
w_T= 
2\tau (G_M^*)^2
+\frac{(G_E^*)^2+\tau (G_M^*)^2}{1+\tau}
\frac{\tau}{\kappa^2}
\left[(\epsilon+\lambda)^2-\kappa^2\frac{1+\tau}{\tau}\right].
\end{equation}

\end{document}